\documentclass[12pt]{article}

\begin{document}

\begin{titlepage}

\begin{flushright}
IUHET-473\\
\end{flushright}
\vskip 2.5cm

\begin{center}
{\Large \bf Gauge Invariance and the Pauli-Villars Regulator in
Lorentz- and CPT-Violating Electrodynamics}
\end{center}

\vspace{1ex}

\begin{center}
{\large B. Altschul\footnote{{\tt baltschu@indiana.edu}}}

\vspace{5mm}
{\sl Department of Physics} \\
{\sl Indiana University} \\
{\sl Bloomington, IN 47405 USA} \\

\end{center}

\vspace{2.5ex}

\medskip

\centerline {\bf Abstract}

\bigskip

We examine the nonperturbative structure of the radiatively induced Chern-Simons
term in a Lorentz- and CPT-violating modification of QED. Although the
coefficient of the induced Chern-Simons term is in general undetermined, the
nonperturbative theory appears to generate a definite value. However, the
CPT-even radiative corrections in this same formulation of the
theory generally break gauge invariance. We show that gauge invariance
may yet
be preserved through the use of a Pauli-Villars regulator, and, contrary to
earlier expectations, this regulator does not necessarily give rise to a
vanishing Chern-Simons term. Instead, two possible values of the Chern-Simons
coefficient are allowed, one zero and one nonzero. This formulation of the
theory
therefore allows the coefficient to vanish naturally, in agreement with
experimental observations.

\bigskip

\end{titlepage}

\newpage

One of the most interesting terms that arises in the study of Lorentz- and
CPT-violating corrections to the standard model
action~\cite{ref-kost1,ref-kost2,ref-kost3} is the electromagnetic Chern-Simons
term, with Lagrange density ${\cal L}_{CS}=\frac{1}{2}(k_{CS})_{\mu}
\epsilon^{\mu\alpha\beta\gamma}
F_{\alpha\beta}A_{\gamma}$~\cite{ref-jackiw2,ref-schonfeld,ref-carroll1}.
${\cal L}_{CS}$ is not gauge invariant; it changes by a total derivative under
a gauge transformation. However, the associated action is gauge invariant, and
it generates a modified theory in which photons of different polarizations
propagate at different speeds. The resulting birefringence has been searched for
in the light from distant
galaxies~\cite{ref-carroll1,ref-carroll2,ref-goldhaber}, but no such effect has
been seen. This constrains $k_{CS}$ to effectively vanish.

It is interesting to ask whether the Chern-Simons term may be generated as a
radiative correction and, if so, whether the physical term's effectively
vanishing value can be reconciled with the possibility of nonzero Lorentz
violation in other sectors of the theory.
The relevant Lagrange density for studying these questions is~\cite{ref-kost4}
\begin{equation}
\label{eq-L}
{\cal L}=-\frac{1}{4}F^{\mu\nu}F_{\mu\nu}+\bar{\psi}(i\!\!\not\!\partial-m-
e\!\!\not\!\!A\,-\!\not\!b\gamma_{5})\psi.
\end{equation}
This theory has the potential to induce a finite
radiatively-generated Chern-Simons
term, with $\Delta k_{CS}$ proportional to $b$. However, the coefficient of
proportionality depends upon the
regularization~\cite{ref-jackiw1,ref-victoria1,ref-victoria2}.
If the regulator used enforces the gauge invariance of the
induced Lagrange density, then $\Delta k_{CS}$ must necessarily
vanish~\cite{ref-coleman}; however, this is not particularly
interesting, because it excludes
the existence of a Chern-Simons term {\em a priori}. Other regulators lead
to different values of $\Delta k_{CS}$, and through a suitable choice, any
coefficient of proportionality between the two may be found. This ambiguity
has been extensively studied, and several potentially interesting values of
$\Delta k_{CS}$ have been
identified~\cite{ref-victoria1,ref-chung1,ref-chung2,ref-chen,
ref-chung3,ref-andrianov}.
Most notably, if the
theory is defined nonperturbatively in $b$, then the ambiguity is lessened, and
a single value appears to be
preferred~\cite{ref-jackiw1,ref-victoria1,ref-chung1}.
However, since this value is nonzero, there is a potential for conflict with the
observed vanishing of $k_{CS}$.

We shall continue the analysis of the nonperturbatively-defined theory.
While the induced Chern-Simons term itself has been extensively studied,
the higher-order, CPT-even corrections to this theory have
largely been neglected. In a nonperturbatively defined theory, the terms of all
orders in $b$ are tied together, and so information about the CPT-even terms
may help clarify the structure of the CPT-odd Chern-Simons term.
We have previously demonstrated~\cite{ref-altschul} that
the ${\cal O}(b^{2})$ terms in the photon self-energy may violate the Ward
identity that enforces the transversality of the vacuum polarization---$p_{\mu}
\Pi^{\mu\nu}(p)=0$; however, our calculation was limited to
the case of $m=0$. We shall demonstrate here that the same result holds in the
opposite limit, when $|b^{2}|\ll m^{2}$ (which represents the physical regime
for all
electrically charged elementary particles). This failure of transversality could
represent a significant problem for the nonperturbative formulation of the
theory. However, this difficulty may be overcome through the use of a
Pauli-Villars regulator. Introducing such a regulator changes the nature of
the ambiguity in the Chern-Simons term. When the Ward-identity-violating terms
are eliminated, we are left with a theory in which the coefficient
$\Delta k_{CS}$ of the induced Chern Simons
term is not in any way forced to vanish, yet it still may
vanish quite naturally. This may represent a resolution of the physical paradox
described above.

We shall calculate the ${\cal O}(b^{2})$ contribution to the zero-momentum
photon self-energy, $\Pi^{\mu\nu}(p=0)$, under the assumption that
$\Pi^{\mu\nu}(p)$ may be expanded as a power series in $b$. However, we do not
expect this assumption to be generically valid. The exact fermion propagator,
\begin{equation}
S(l)=\frac{i}{\!\not l-m\,-\!\not\!b\gamma_{5}},
\end{equation}
may be rationalized to obtain~\cite{ref-victoria1,ref-chung1}
\begin{equation}
\label{eq-propagator}
S(l)=i\frac{(\!\not l+m\,-\!\not\!b\gamma_{5})(l^{2}-m^{2}-b^{2}+[\!\not l,
\!\not\!b\,]\gamma_{5})}{(l^{2}-m^{2}-b^{2})^{2}+4[l^{2}b^{2}-(l\cdot b)^{2}]}.
\end{equation}
At $l=0$, the denominator of the rationalized propagator becomes
$(m^{2}+b^{2})^{2}$. The square root $|m^{2}+b^{2}|$ of this expression arises
in the calculation of $\Pi^{\mu\nu}(p=0)$, and the absolute value leads to
behavior that is nonanalytic in $b$. The $b$-odd portion of the self-energy
(which is just the Chern-Simons term) has different forms for $-b^{2}\leq m^{2}
\neq0$, $-b^{2}<m^{2}\neq0$, and $m=0$.
For the $b$-even terms, we also expect the power-series
representation of the self-energy to break down at $b^{2}=-m^{2}$;
moreover, there may exist other thresholds for nonanalytic behavior as well.
We shall therefore restrict our attention to the regime in which
$|b^{2}|\ll m^{2}$. Since large values of $b^{2}$ are excluded for most
physical particles, this is a reasonable restriction.

To control the high-energy behavior of our theory, we shall use a Pauli-Villars
regulator.
By utilizing this method of regulation, we may deal with the usual divergences
that arise in the photon self-energy [at ${\cal O}(b^{0})$], as well as the
Ward-identity-violating terms that appear at ${\cal O}(b^{2})$. The use of a
single regulator at all orders in $b$ is necessary, because in the
nonperturbative formalism, there is actually only a single Feynman diagram that
contributes to the photon self-energy at ${\cal O}(e^{2})$. This diagram is the
usual QED vacuum polarization, but with the usual fermion propagator replaced
by the $b$-exact $S(l)$. We shall use a symmetric integration prescription
in our calculation, because this is appropriate at ${\cal O}(b^{0})$, and it
preserves the expected
transformation properties of the
loop integral.
(Moreover, without
symmetric integration, additional ambiguities in $\Delta k_{CS}$ would exist.)

The Pauli-Villars regularization of the photon self-energy involves the
introduction of a fictitious species of heavy fermions, whose contribution to
the self-energy is subtractive. This subtraction renders the Lorentz-invariant
part of the self-energy finite and preserves its gauge invariance. Previously,
it has been believed that the same subtraction will cause the ${\cal O}(b)$
Chern-Simons term to vanish. However, we shall show that this is not necessarily
the case.

The fictitious Pauli-Villars particles have the exact propagator
\begin{equation}
S_{M}(l)=\frac{i}{\!\not l-M\,-\!\not\!b_{M}\gamma_{5}},
\end{equation}
where $M$ is the particles' large mass, and $b_{M}$ is the Lorentz-violating
coefficient appropriate to these fictitious particles. Previous analyses have
assumed that $b_{M}=b$, and indeed this must be the case in a
perturbatively defined theory, in which the $b$ interaction is treated as a
vertex. However, in the nonperturbative formulation, this restriction is absent,
and $b$ and $b_{M}$ exist as distinct objects.
Depending upon what properties we demand for the self-energy, $b_{M}$ may take
on different values. Even if we place quite strong (yet quite natural)
conditions on the form of $\Pi^{\mu\nu}(p)$, we are still left with a discrete
freedom in our choice of $b_{M}$; either $b_{M}=b$ or
$b_{M}=-b$ will suffice.

In fact, the weakest condition that we may place upon $b_{M}$ is no condition at
all. If we choose not to worry about the failure of the Ward identity at
${\cal O}(b^{2})$, then the only restriction on the Pauli-Villars propagator
is that it must cancel the divergences at ${\cal O}(b^{0})$. In that case,
any value of $b_{M}$ is acceptable. We might be tempted to choose a vanishing
$b_{M}$, which would produce a formulation of the theory equivalent to that
used in~\cite{ref-jackiw1}.
However, since Lorentz violation is already present in the theory, there is
no reason to prefer this value. Similarly, the choice $b_{M}=b$,
while aesthetically pleasing, has nothing special to recommend it.
We are left with an arbitrary $b_{M}$, which
will generate an arbitrary Chern-Simons term. Moreover, since $b$ and $b_{M}$
need not point in the same direction, the coefficient $\Delta k_{CS}$ of the
induced Chern-Simons term need not even be parallel to $b$; this is an even
greater degree or arbitrariness in $\Delta k_{CS}$ than has been found
previously.

The above scenario illustrates an interesting point. In the theory as we have
described it, there can exist a radiatively-induced Chern-Simons term even if
$b$ vanishes, because the regulator involves the Lorentz-violating parameter
$b_{M}$. In a general
Lorentz-violating quantum field theory, the Lagrange density need
not be the only source of the violation. The regularization prescription can
also break Lorentz invariance, and the Lorentz violation of the regulator need
not be determined by the behavior of any terms in the unrenormalized action.
It may seem somewhat unnatural for the regulator to become an additional source
of Lorentz violation. However,
it is well established that fictitious Pauli-Villars fermions may generate
radiative corrections that are qualitatively different from those generated by
a theory's real fermions.
For example, in massless QED, with a Pauli-Villars regulator, it is the
fictitious heavy particles which are solely responsible for the anomalous
nonconservation of the axial vector and dilation currents.
However, the specific scenario presented
in the previous paragraph is probably unrealistic, because
we had to abandon the Ward identity in order to avoid placing any
restrictions on $b_{M}$.

The key to developing a more sophisticated theory---one in which gauge
invariance holds at all orders---is the determination of the ${\cal O}(b^{2})$
part of the self-energy. Since we are only interested in the situation in which
we may expand $\Pi^{\mu\nu}(p)$ as a power series in $b$, we may use the
approximation scheme developed in~\cite{ref-jackiw1} to expand the
exact propagator to second order in $b$. We find
\begin{equation}
\label{eq-Sappr}
S(l)\approx\frac{i}{\!\not l-m}+\frac{i}{\!\not l-m}\left(-i
\!\not\!b\gamma_{5}\right)\frac{i}{\!\not l-m}+\frac{i}{\!\not l-m}
\left(-i\!\not\!b\gamma_{5}\right)\frac{i}{\!\not l-m}
\left(-i\!\not\!b\gamma_{5}\right)\frac{i}{\!\not l-m}.
\end{equation}
Naively, this might appear equivalent to treating the $b$ term as an interaction
vertex. However, there are subtle differences that characterize our
nonperturbative approach; since there is only a single diagram, there is a
complex interplay between terms at different orders in $b$.

The one-loop photon self-energy is
\begin{equation}
\label{eq-Pi}
\Pi^{\mu\nu}(p)=-ie^{2}\int\frac{d^{4}k}{(2\pi)^{4}}{\rm tr}\left\{
\gamma^{\mu}S(k)\gamma^{\nu}S(k+p)\right\}+\Pi_{M}^{\mu\nu}(p).
\end{equation}
The first term on the right-hand side of (\ref{eq-Pi}) is the ``fermion part"
of the total self-energy, while $\Pi_{M}^{\mu\nu}(p)$
represents the ``Pauli-Villars part." We see that the
${\cal O}(b^{2})$ contribution to $\Pi^{\mu\nu}(p)$ has a potential logarithmic
divergence. However, if we
were to improve the approximation (\ref{eq-Sappr}) to obtain
higher-order corrections in $b$, then the corresponding
higher-order corrections to $\Pi^{\mu\nu}(p)$ will involve manifestly
convergent integrals. Similarly, any ${\cal O}(b^{2})$
terms involving positive powers of $p$ will also be power-counting
finite. These finite terms
must automatically obey the Ward identity, so we expect that by calculating
the second-order portion of $\Pi^{\mu\nu}(p=0)$, we should be isolating all
the possible sources of Ward identity violation.

In fact, the ${\cal O}(b^{2})$ contribution to the fermion part of
the self-energy is
\begin{eqnarray}
\Pi^{\mu\nu}_{b^{2}}(0) & = & -ie^{2}\int\frac{d^{4}k}{(2\pi)^{4}}\,{\rm tr}
\left\{\gamma^{\mu}\frac{i}{\!\not k-m}
\left(-i\!\not\!b\gamma_{5}\right)\frac{i}{\!\not k-m}
\gamma^{\nu}\frac{i}{\!\not k-m}
\left(-i\!\not\!b\gamma_{5}\right)\frac{i}{\!\not k-m}\right. \nonumber\\
& & \left.\left[\gamma^{\mu}\frac{i}{\!\not k-m}\gamma^{\nu}\frac{i}{\!\not k-m}
\left(-i \!\not\!b \gamma_{5}\right)\frac{i}{\!\not k-m}
\left(-i\!\not\!b\gamma_{5}\right)\frac{i}{\!\not k-m}
+(\mu\leftrightarrow\nu)\right]\right\}. \\
& = & ie^{2}\int\frac{d^{4}k}{(2\pi)^{4}}\frac{1}{\left(k^{2}-m^{2}\right)^{4}}
\,{\rm tr}\left\{\gamma^{\mu}(\!\not k+m)\!\not\! b\,(\!\not k-m)\gamma^{\nu}
(\!\not k-m)\!\not\! b\,(\!\not k+m)\right. \nonumber\\
& & +\left.\left[
\gamma^{\mu}(\!\not k+m)\gamma^{\nu}(\!\not k+m)\!\not\! b\,(\!\not k
-m)\!\not\!b\, (\!\not k+m)+(\mu\leftrightarrow\nu)\right]\right\}
\end{eqnarray}
The only Lorentz structures that may be present in $\Pi^{\mu\nu}_{b^{2}}(0)$ are
$b^{\mu}b^{\nu}$ and $g^{\mu\nu}b^{2}$, so it is simplest to evaluate the
self-energy by calculating $g_{\mu\nu}\Pi^{\mu\nu}_{b^{2}}(0)$ and
$\frac{b_{\mu}b_{\nu}}{b^{2}}\Pi^{\mu\nu}_{b^{2}}(0)$. These calculations
are fairly straightforward. For $g_{\mu\nu}\Pi^{\mu\nu}_{b^{2}}(0)$, we have
\begin{equation}
g_{\mu\nu}\Pi^{\mu\nu}_{b^{2}}(0)=-ie^{2}\int\frac{d^{4}k}{(2\pi)^{4}}
\frac{{\rm tr}\{[2(\!\not k\!\not\! b\!\not k-m^{2}\!\not\! b)+4(\!\not k-2m)
(\!\not k+m)\!\not\! b](\!\not k-m)\!\not\! b(\!\not k+m)\}}
{\left(k^{2}-m^{2}\right)^{4}}.
\end{equation}
The ${\cal O}(k^{4})$ terms in the numerator cancel under symmetric integration,
and we are left with the finite integral
\begin{eqnarray}
g_{\mu\nu}\Pi^{\mu\nu}_{b^{2}}(0) & = & 4ie^{2}\int\frac{d^{4}k}{(2\pi)^{4}}
\frac{16m^{2}[2(b\cdot k)^{2}-b^{2}k^{2}]-10m^{4}b^{2}}
{\left(k^{2}-m^{2}\right)^{4}} \\
& = & -\frac{e^{2}b^{2}}{4\pi^{2}}.
\end{eqnarray}
Similarly, we also have
\begin{eqnarray}
\frac{b_{\mu}b_{\nu}}{b^{2}}\Pi^{\mu\nu}_{b^{2}}(0) & = & \frac{ie^{2}}{b^{2}}
\int\frac{d^{4}k}
{(2\pi)^{4}}\frac{{\rm tr}\{\!\not\! b(\!\not k +m)\!\not\! b(3\!\!\!\not k+m)
\!\not\! b(\!\not k-m)\!\not\! b(\!\not k+m)\}}
{\left(k^{2}-m^{2}\right)^{4}} \\
& = & -4ie^{2}\int\frac{d^{4}k}{(2\pi)^{4}}\frac{2m^{2}b^{2}k^{2}+m^{4}b^{2}}
{\left(k^{2}-m^{2}\right)^{4}} \\
& = & -\frac{e^{2}b^{2}}{8\pi^{2}},
\end{eqnarray}
where the logarithmic divergences again have canceled out.

From these results, it follows that
\begin{equation}
\label{eq-Pib2}
\Pi^{\mu\nu}_{b^{2}}(0) = -\frac{e^{2}}{24\pi^{2}}\left(2b^{\mu}b^{\nu}+
g^{\mu\nu}b^{2}\right).
\end{equation}
This is the same result that we found previously in the $m=0$
case~\cite{ref-altschul}. Since $p_{\mu}(2b^{\mu}b^{\nu}+g^{\mu\nu}b^{2})\neq0$,
there is the potential for a violation of gauge invariance.
However, since we are using
a Pauli-Villars regularization, there is also a contribution from
$\Pi^{\mu\nu}_{M}(p)$ that is second order in $b_{M}$. This
contribution is precisely
$\frac{e^{2}}{24\pi^{2}}\left(2b_{M}^{\mu}b_{M}^{\nu}+g^{\mu\nu}b_{M}^{2}
\right)$. We are now in a position to place a natural condition on $b_{M}$;
if transversality is to
be maintained, the ${\cal O}(b^{2})$ and ${\cal O}(b_{M}^{2})$ contributions
must cancel. This cancellation occurs exactly if $b_{M}=\pm b$. Therefore, we
are left
with a binary choice between two theories, characterized by different relative
signs of $b$ and $b_{M}$, and these theories exhibit different values of the
radiatively induced Chern-Simons coefficient. If $b_{M}=b$, then $\Delta k_{CS}$
vanishes, in accordance with earlier expectations. However, if $b_{M}=-b$, then
\begin{equation}
\Delta k_{CS}^{\mu}=\frac{3e^{2}}{8\pi^{2}}b^{\mu};
\end{equation}
this is twice the value found in~\cite{ref-jackiw1,ref-victoria1,ref-chung1},
since there are equal contributions from $b$ and $b_{M}$.

This is a novel situation. Unambiguously finite yet undetermined radiative
corrections generally involve continuously variable parameters that
describe the arbitrariness of the results~\cite{ref-jackiw3}, and indeed,
in more general formulations of the theory we are discussing, there is a
continuous ambiguity in the induced Chern-Simons term. However, in
this instance, only a discrete set of values for $\Delta k_{CS}$ is
allowed. Since zero is one of these allowed values, this formulation allows for
the existence of a gauge-invariant Lagrange density, but it does not force the
density to be so invariant. The Pauli-Villars formulation therefore allows the
theory to possess a naturally vanishing induced Chern-Simons term without
externally enforcing the gauge invariance of the Lagrange density. Moreover,
since the Pauli-Villars regulator is covariant, it should be possible to
generalize our calculations to occur in a weakly curved
spacetime~\cite{ref-kost5}.

As we remarked in~\cite{ref-altschul}, the ${\cal O}(b)$ and ${\cal O}(b^{2})$
terms in the photon self-energy both violate gauge invariance in the
strongest fashions that are allowed by their tensor structures. The Chern-Simons
term violates the gauge invariance of the Lagrange density; however, because the
induced density
necessarily involves the antisymmetric $\epsilon$-tensor, the Ward identity is
preserved and the integrated action remains gauge invariant.
$\Pi^{\mu\nu}_{b^{2}}(p)$ possesses no such special structure, and so it
violates the Ward identity explicitly. In order to ensure gauge invariance, we
have therefore
been forced to place a very strong condition on the tensor structure of
the second-order correction to the self-energy; we have insisted that it must
vanish at $p=0$. Since there is more than one way to enforce this
condition, we have found more than one possible value for $\Delta k_{CS}$. This
is the origin of the discrete ambiguity in this formulation of the theory.

We may now ask what may happen if the mass scale $m^{2}$ is not large compared
with the Lorentz-violation scale $|b^{2}|$. When the real fermions are massless,
their ${\cal O}(b^{2})$ contribution to $\Pi^{\mu\nu}(p)$ is exactly the
$\Pi^{\mu\nu}_{b^{2}}(0)$ of equation (\ref{eq-Pib2}). In this instance, as
in the $|b^{2}|\ll m^{2}$ case, the use of a power series expansion to determine
the second-order portion
of the photon self-energy is justified, so the similarity
of the results is somewhat unsurprising.
If a Pauli-Villars regulator is
used in this case, the Ward identity may be preserved, but the Chern-Simons term
cannot be made to vanish. The choice is between $\Delta k_{CS}^{\mu}=
\frac{e^{2}}{8\pi^{2}}b^{\mu}$ and $\Delta k_{CS}^{\mu}=-\frac{e^{2}}
{4\pi^{2}}b^{\mu}$.
So this formulation of the theory
does not allow for $\Delta k_{CS}$ to vanish naturally if there
are massless fermions coupled to the gauge field.

When a power series expansion in $b$ is not justified, the gauge noninvariant
portions of the self-energy must be calculated directly from the
nonperturbative propagator (\ref{eq-propagator}). Only if the result is of the
form $A(b^{2})b^{\mu}b^{\nu}+B(b^{2})g^{\mu\nu}b^{2}$, with
$A(b^{2})=2B(b^{2})\leq 0$, will the Pauli-Villars regulator be capable of
restoring the Ward identity. If we do choose to
use a Pauli-Villars regulator and also insist upon gauge invariance
as a requirement of our theory, then any breakdowns of the Ward identity when
$|b^{2}|\not\ll m^{2}$ can be interpreted as
giving restrictions on the allowed values
of $b$. Applied to the standard model, this could imply that the scale of
$b$-generated Lorentz violation should be smaller than the smallest
nonzero fermionic mass scale present. However, this is all merely conjectural,
since we do not know the general form taken by the self-energy when $|b^{2}|
\not\ll m^{2}$, and because of the
substantial additional complexity of the standard model.

This work has demonstrated that there exist new, unexpected subtleties in the
structure of the radiatively-induced Chern-Simons term.
We have identified new types of
ambiguities in the value of the Chern-Simons coefficient; these ambiguities are
derived from the use of a Pauli-Villars regulator and
a nonperturbative formulation of the theory. If we do not insist
that the Ward identity hold at second order, then the coefficient is completely
arbitrary and is not even constrained to be parallel to $b$. However, if, more
reasonably, we do insist on transversality of the vacuum polarization
at higher orders,
then there is a binary ambiguity in the value of the induced coefficient.
If $m=0$, either of two nonzero values for $\Delta k_{CS}$ is allowed.
However, if $|b^{2}|\ll m^{2}$, then $\Delta k_{CS}=0$ is allowed, in addition
to a particular nonzero value. This may provide a satisfactory explanation for
why the electromagnetic Chern-Simons coefficient vanishes in nature.

\section*{Acknowledgments}
The author is grateful to V. A. Kosteleck\'{y}, M. P\'{e}rez-Victoria, and
R. Jackiw for many helpful discussions.
This work is supported in part by funds provided by the U. S.
Department of Energy (D.O.E.) under cooperative research agreement
DE-FG02-91ER40661.

\end{document}